\documentclass[10pt,conference]{IEEEtran}

\usepackage{times}
\usepackage[cmex10]{amsmath}
\usepackage{amssymb}
\usepackage{epsfig,verbatim}

\newcommand{\eqnlabel}[1]{\label{eqn:#1}}
\newcommand{\eqnref}[1]{(\ref{eqn:#1})}

\providecommand{\xv}{\mathbf{x}} 
\providecommand{\yv}{\mathbf{y}}

\providecommand{\Xc}{{\cal X}}
\providecommand{\Yc}{{\cal Y}}

\providecommand{\Wc}{{\cal W}}

\newtheorem{MyTheorem}{Theorem}
\newcommand{\thmlabel}[1]{\label{thm:#1}}
\newcommand{\thmref}[1]{\ref{thm:#1}}

\newtheorem{MyRemark}{Remark}
\newcommand{\remarklabel}[1]{\label{thm:#1}}

\title{Short Message Noisy Network Coding with \\  Rate Splitting}
\begin{document}




\author{
\IEEEauthorblockN{Ivana Mari\'c
              and Dennis Hui}
\IEEEauthorblockA{Ericsson Research\\
              200 Holger Way, San Jose, CA, USA\\
              Email: \{ivana.maric, dennis.hui\}@ericsson.com}
}

\maketitle

\date{}

\begin{abstract}
For noisy network with multiple relays, short message noisy network coding with rate splitting (SNNC-RS) scheme is presented.
It has recently been shown by Hou and Kramer that mixed cooperative strategies in which relays in favorable positions decode-and-forward (DF) and the rest quantize via short message noisy network coding (SNNC) can outperform existing cooperative strategies (e.g., noisy network coding (NNC) and decode-and-forward). A drawback of such schemes is  that forwarding of quantized signals at SNNC relays introduces interference to the rest of the relays. Our proposed relaying scheme has a capability to reduce such  interference and thereby improve the rate performance.   In the proposed scheme, superposition coding is incorporated into SNNC encoding to enable partial interference cancellation at DF relays.   The achievable rate with proposed scheme is derived for the discrete two-relay network and evaluated in the Gaussian case where  gains over the rate achievable without rate splitting are demonstrated.

\end{abstract}
\section{Introduction}
General capacity achieving encoding strategies for wireless networks are unknown. 
Two coding  strategies, {\it noisy network coding} (NNC) \cite{Avestimehr2011}, \cite{LimKimElGamal2011} and {\it short message NNC} (SNNC) \cite{HouKramer2013} have recently been shown to outperform other cooperative strategies for multicast networks (e.g., decode-and-forward (DF)) \cite{LimKimElGamal2011},\cite{HouKramer2013}. Both NNC and SNNC are  based on compression at relays, first used in compress-and-forward (CF) \cite{CoverElGamal79}.
However, in networks with many relays receiving  signals of different strengths, constraining all relays to perform the same cooperative scheme may not be optimal. 
Relays in favorable positions can perform DF thereby removing the noise otherwise partially propagated via compression based schemes. 
On the other hand, decoding requirement at a DF relay can severly limit the transmission rate if the links over which the relay is receiving  are weak.
It has recently been shown by Hou and Kramer that  {\it mixed} cooperative strategies that allow some relays to perform  DF while the rest deploy SNNC,  can {\it outperform} strategies in which all relays use the same relaying scheme (e.g., DF, CF, NNC or SNNC) \cite{HouKramer2013}.
 NNC and SNNC  can achieve the same transmission rate \cite{HouKramer2013}; the main difference between them is that in SNNC, the source sends independent "short" messages in each block, whereas in NNC, the same message of {\it higher} encoding rate is repeatedly sent over many blocks.
The lower encoding  rate of short messages  allows relays which have strong received signals to perform DF, while other relays  still deploy SNNC \cite{KramerHou2011}, thus enabling the use of mixed strategies.

In this paper, we   present an encoding scheme that improves performance achievable with the mixed cooperative strategies analyzed in \cite{HouKramer2013}.  
To motivate our approach, consider a multihop relay network  in which all relays use DF, shown in Fig.~1.  A relay decodes a message based on signals received from all "upstream" nodes. At the same time, although all relays simultaneously transmit, "downstream" relays do not cause interference to the upstream relays because the latter ones know the messages sent by the downstream relays and can cancel created interference \cite{KramerGastparGupta2005}. 
The limitation of this scheme is that, if the received signal is weak at even one DF relay,  the end-to-end rate will be decreased; relays with weak received signals should  compress their signals. However, when some of the relays compress,  interference cancellation at DF relays is {\it no longer possible} because DF nodes do not know the compressed signals.  This can  decrease the rate at DF relays and thus the overall performance. To overcome this drawback of  mixed cooperative strategies, we present the {\it SNNC with rate splitting} scheme that allows DF relays to partially decode interference created by SNNC relays. The idea  is to incorporate superposition coding \cite{Cover} into  SNNC encoding. 
In particular, in the proposed scheme, a relay performing SNNC will, once it determines the compression index it wants to send,   split  the compression index into two indexes each of a lower rate and  use superoposition coding to encode them. The lower encoding rate enables  DF relays  to decode one part of the quantization index and hence cancel part of interference. This in turn increases the end-to-end rate.

The main contributions of this paper are the following:

\begin{figure}[t]
\centering
\includegraphics[height=.3in, width=3.5in]{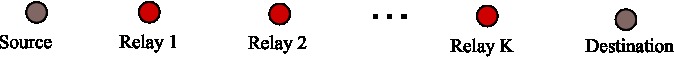}
\caption{Single source-destination network with multiple relays. Relays transmit sequentially using DF.}
     \label{Fig1}
\end{figure}
\begin{enumerate}
\item A novel  relaying strategy is proposed and analyzed. 
\item It is demonstrated that, in a single-relay channel \cite{CoverElGamal79}, rate splitting does not reduce SNNC rate. Note that in this channel, rate splitting cannot bring gains because there are no multiple relays deploying different  schemes.
\item  For the general discrete memoryless two-relay network (shown in Fig.~\ref{Fig2}), an achievable rate of the proposed cooperative strategy  is derived. 
\item The obtained rate is evaluated for Gaussian two-relay networks; rate gains of the proposed SNNC-RS coding scheme compared to other coding schemes are demonstrated. 
\end{enumerate}

\section{Main Result} 
\begin{figure}[t]
\centering
\includegraphics[height=1.2in, width=2.8in]{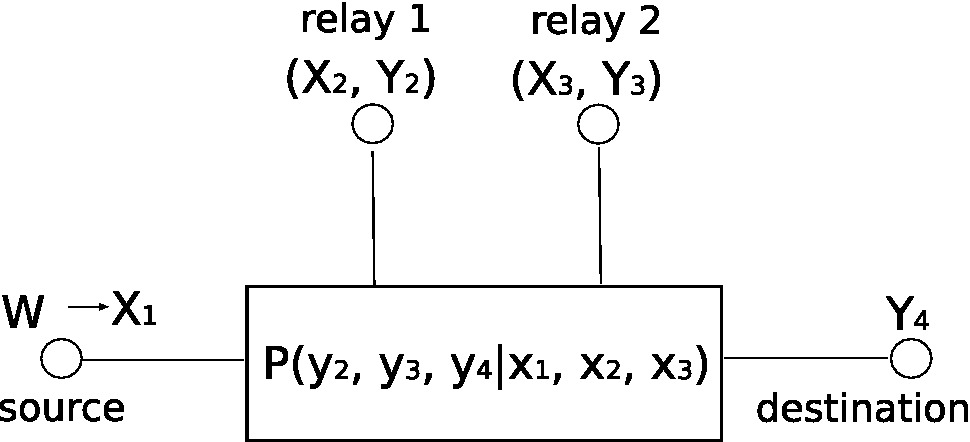}
\caption{Two-Relay Channel Model.}
     \label{Fig2}
\end{figure}
We denote random variables with upper case letters and their realizations with the corresponding lower case letters. 
We drop subscripts of probability distributions if the arguments of the distributions are lower case versions of the random variables.

Consider a discrete memoryless two-relay channel shown in Fig.~\ref{Fig2}. 
This is the smallest network that captures gains from our proposed scheme.
The source wishes to send message $W$ from the message set $\Wc$ to the destination node $4$. The channel is described by the conditional probabilities $P(y_2,y_3,y_4|x_1,x_2,x_3)$ where $x_i \in {\cal X}_i$, $i=1,2,3$ and $y_j \in {\cal Y}_j$, $j=2,3,4$ and ${\Xc}_i$ and ${\Yc}_j$ are respective input and output alphabets at nodes $i$ and $j$. 

A $(R, n)$ code for the two-relay network consists of the message set $\Wc = \{1, \ldots , 2^{nR}\}$,  
encoding functions at the source $X_1^n = f(W)$, and at  the relays $X_{k,i} = f_{k,i} (Y^ {i-1}_k ), k=2,3$ 
and the decoding function $\hat W = g(Y_4^n )$. The average error probability
of the code is given by $P_e = P[ \hat W \neq W ] $. A
rate $R$ is achievable if, for any $ \epsilon> 0$, there exists,
for a sufficiently large $n$, a code $(R, n)$ such that $P_e < \epsilon$.
The capacity  is the supremum of all achievable $R$.

We analyze an encoding strategy in which relay $1$ deploys DF, and relay $2$ deploys SNNC-RS. The code construction, encoding and decoding are specified in the Appendix.  Quantization at relay $2$ is performed as in SNNC; the difference between SNNC-RS and SNNC schemes is in the addition of superposition coding in SNNC-RS which is used to encode the quantization index at relay $2$. Specifically, instead of using a single codebook to encode
the quantization index (say at rate $R_3$), relay $2$ splits the quantization index into two indexes and encodes them via superposition coding using two codebooks at rates $R_{30}$ and
$R_{31}$, such that $R_3 = R_{30} + R_{31}$. Each codebook thus has a
lower rate than $R_3$.  The lower encoding rate (than $R_3$) used in SNNC-RS enables relay $1$ to, in addition to decoding the source message $W$, also decode one part of the quantization index sent by relay $2$ thereby reducing observed interference.  We consider joint decoding at relay $1$. The destination uses backward decoding to decode the source message and the quantization index.
Before analyzing the proposed scheme for the two-relay channel, we show the following important property of the SNNC-RS scheme for the single-relay channel \cite{CoverElGamal79}.
\begin{MyTheorem}
Rate splitting at the relay does not decrease the SNNC rate in the single-relay channel.
\end{MyTheorem}
The proof can be found in \cite{MaricArxiv2014}.

 We have the following result for the two-relay channel.

\begin{MyTheorem} \thmlabel{MainResult}
{\it (Joint decoding at relay 1)}
The achievable rate with coding scheme in which relay 1 performs DF and SNNC-RS and joint decoding at the DF relay in the two-relay channel satisfies:
\begin{align}
& R < I(X_1;Y_2|X_2 X_{30}) \eqnlabel{jd1} \\
&R < I(\bar X; \hat Y_3 Y_4|X_{30} X_{31}) \eqnlabel{jd2} \\
& R < I (\bar X X_{30} X_{31}; Y_4) -  I(\hat Y_3; Y_3|\bar X X_{30} X_{31} Y_4)    \eqnlabel{jd3} \\
&R <I ( X_{31}; Y_4| \bar X X_{30}) - I(\hat Y_3; Y_3|\bar X X_{30} X_{31} Y_4)   \nonumber \\
& \qquad + I(X_1 X_{30}; Y_2|X_2) \eqnlabel{jd9} \\
&2R <  I (\bar X X_{31}; Y_4| X_{30}) - I(\hat Y_3; Y_3|\bar X X_{30} X_{31} Y_4)  \nonumber \\
& \qquad   + I(X_1 X_{30}; Y_2|X_2) \eqnlabel{2R} 
\end{align}
for  any joint distributions that factors as
\begin{equation}\eqnlabel{pmf}
P(x_1 x_2) P(x_{30}x_{31}) P (\hat y_3|x_{30}x_{31} y_3) P(y_2 y_3 y_4|x_1 x_2 x_{30}x_{31})
\end{equation}
and
where we introduced the notation
$\bar X =(X_1,X_2)$.
\end{MyTheorem}
\begin{IEEEproof}
The proof outline  is given in the Appendix.
\end{IEEEproof}

\begin{MyRemark}
In the special case of no rate-splitting , i.e., for $X_{30}=\emptyset, X_{31}=X_3$, the rate of Thm.~\thmref{MainResult} is the DF-SNNC rate without rate splitting given by \cite[Eq. (56)]{HouKramer2013}. For this case,  \eqnref{jd9}-\eqnref{2R} are loose and  \eqnref{jd1}-\eqnref{jd3} coincide with the achievable rate   \cite[Eq. (56)]{HouKramer2013} given by 
\begin{align}
& R < I(X_1;Y_2|X_2) \nonumber \\
&R < I(X_1 X_2; \hat Y_3 Y_4|X_3) \nonumber \\
& R < I (X_1 X_2 X_3; Y_4) -  I(\hat Y_3; Y_3| X_1 X_2 X_3 Y_4)    \eqnlabel{K1} 
\end{align}
for any joint distribution factors that factors as \eqnref{pmf}.
\end{MyRemark}
\begin{MyRemark}
When condition 
\begin{equation}
I(X_{30}; Y_4| \bar X) < I(X_{30}; Y_2| \bar X) \eqnlabel{remark1}
\end{equation}
is satisfied,  relay 1 can decode the full quantization index sent by relay 2. 
\end{MyRemark}
\begin{MyRemark} \remarklabel{Remark2}
When
\begin{equation}
I(X_{30}; Y_4) < I(X_{30}; Y_2| X_2)     \eqnlabel{simplecondition}
\end{equation}
bound \eqnref{2R} is loose compared to the sum of the rates given by \eqnref{jd1} and \eqnref{jd3}.
\end{MyRemark}
\begin{MyRemark}
When
\begin{equation}
I(\bar X X_{30}; Y_4) < I(X_1 X_{30}; Y_2|  X_2 ) \eqnlabel{simplecondition2}
\end{equation}
bound \eqnref{jd9} is loose compared to \eqnref{jd3}.
\end{MyRemark}

In the case that the relay $1$ uses successive decoding whereby it first decodes the part of the quantization index and then the source message, the corresponding achievable region with mixed DF and SNNC-RS is given by the next theorem.
\begin{MyTheorem} {\it (Successive decoding at relay $1$)}
The achievable rate with mixed strategy of DF and SNNC-RS and sequential decoding at the DF relay in the two-relay channel satisfies: 
\begin{align}
&R < I(X_1;Y_2|X_2 X_{30}) \nonumber\\
&R < I(\bar X; \hat Y_3 Y_4|X_{30} X_{31}) \nonumber \\
&R< I(\bar X X_{30} X_{31}; Y_4) - I(\hat Y_3; Y_3|\bar X X_{30} X_{31} Y_4) \nonumber\\
&R< I(\bar X X_{31}; Y_4| X_{30} ) -I(\hat Y_3; Y_3|\bar X X_{30} X_{31} Y_4) \nonumber \\
& \qquad  + I( X_{30}; Y_2|X_2) \nonumber\\
&   I(X_{31}; Y_4|\bar X X_{30}) +  I(X_{30}; Y_2|X_2) \nonumber \\
& \qquad  -  I(\hat Y_3; Y_3|\bar X X_{30} X_{31} Y_4) >0\eqnlabel{sd5}
\end{align}
for any joint distribution given by \eqnref{pmf}.
\end{MyTheorem}
The proof follows the same steps as the proof for Theorem \thmref{MainResult}.

\begin{figure}[t]
\centering
\includegraphics[scale=.45]{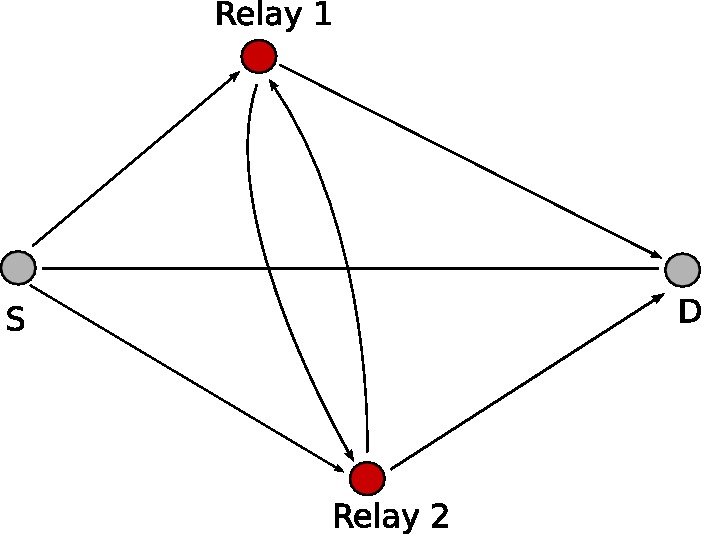}
\caption{Two-Relay Gaussian Channel.}
     \label{Fig3}
\end{figure}

We next evaluate the rate given by Theorem \thmref{MainResult} for Gaussian channels. We then compare it to the rate achieved with the mixed encoding scheme that does not use rate splitting, as well as to the schemes in which both relays decode-and-forward or perform noisy network coding.

\section{Gaussian Channel} \label{Gauss}

We evaluate the obtained rate \eqnref{jd1}-\eqnref{2R}  in Gaussian channels 
\begin{align}
& Y_2 = h_{12} X_1 + h_{32} X_3 + Z_2 \nonumber \\
& Y_3 = h_{13} X_1 + h_{23} X_2 + Z_3 \nonumber \\
& Y_4 = h_{14} X_1 + h_{24} X_2 + h_{34} X_3 + Z_4 \eqnlabel{GaussChannel}
\end{align}
where $Z_i \sim {\cal N}(0,1)$ is additive white Gaussian noise. 
Channel gain from  node $i$ to node $j$ is denoted as $h_{ij}$. 
We assume average power constraint $P_i$ at each node $i$, $=1,2,3$ given by 
$E[X_i^2] \le P_i.$
We choose Gaussian inputs $X_i \sim{\cal N}(0,P_i)$.
We denote $C(x) = \log_2 (1+x).$
Rate bounds \eqnref{jd1}-\eqnref{2R} evaluate to
\begin{align}
&R < C \left( \frac{\bar \beta h_{12}^2 P_1}{1 + \bar \alpha h_{32}^2 P_3} \right) \nonumber \\
&R < C \left(  P_1(\frac{h_{13}^2}{1+ \hat N_3} + h_{14}^2)  + P_2(\frac{h_{23}^2}{1+ \hat N_3} + h_{24}^2) \right. \nonumber \\
& \qquad + 2 \sqrt{\beta P_1 P_2} (\frac{h_{13} h_{23}}{1+ \hat N_3}+ h_{14} h_{24})  \nonumber \\
& \qquad \left. + \frac {\bar \beta P_1 P_2}{1+ \hat N_3}  (h_{13} h_{24} - h_{23} h_{14})^2 \right) \nonumber 
\end{align}
\begin{align}
& R < C \left( h_{14}^2 P_1 + h_{24}^2 P_2 + 2h_{14}h_{24} \sqrt{\beta P_1 P_2} + h_{34}^2 P_3 \right) \nonumber \\
& \qquad - C \left( \frac{1}{\hat N_3} \right)  \nonumber \\
& R < C \left(  \bar \alpha h_{34}^2 P_3 \right)+ C \left( \frac{ \bar \beta h_{12}^2 P_1 + \alpha h_{32}^2 P_3}{1 + \bar \alpha h_{32}^2 P_3} \right)   - C \left( \frac{1}{\hat N_3} \right)  \nonumber \\
& 2R <  C \left( h_{14}^2 P_1 + h_{24}^2 P_2 + 2h_{14}h_{24} \sqrt{\beta P_1 P_2} + \bar \alpha h_{34}^2 P_3 \right) \nonumber \\
& \qquad - C \left( \frac{1}{\hat N_3} \right) + C \left( \frac{ \alpha h_{32}^2 P_3}{1 + \bar \alpha h_{32}^2 P_3} \right) + C \left( \frac{\bar \beta h_{12}^2 P_1}{1+ h_{32}^2 P_3} \right). \eqnlabel{G1}
\end{align} 
\begin{MyRemark}
As pointed out in the previous section, when condition \eqnref{remark1} is met, full decoding of quantization index at relay $1$ is possible. This is easily observed in the considered Gaussian channel \eqnref{GaussChannel} because \eqnref{remark1} evaluates to
\begin{equation} \eqnlabel{full}
h_{32} > h_{34}.
\end{equation}
\end{MyRemark}
\begin{figure}[t]
\centering
\includegraphics[scale=0.45]{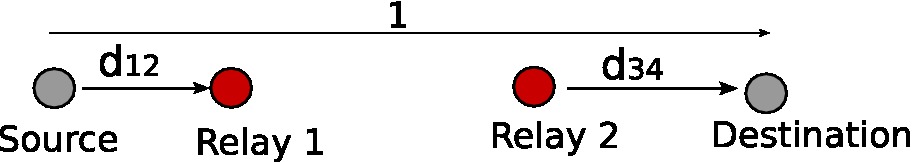}
\caption{Two-Relay Channel Example.}
     \label{Fig5}
\end{figure}

\begin{figure}[t]
\centering
\includegraphics[height=2.8in, width=3.5in]{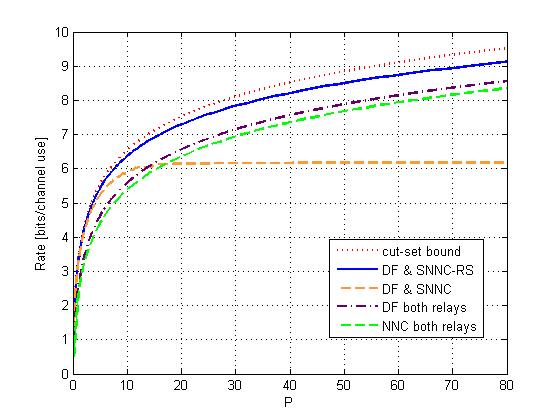}
\caption{Rate comparison of coding schemes for $d_{12} = 0.1$, $d_{34} = 0.05$, $d_{14} = 1$.}
     \label{Fig4}
\end{figure}

Fig.~\ref{Fig4} shows the comparison of the proposed scheme with  the corresponding DF-SNNC scheme without rate splitting \eqnref{K1}, for a network in Fig.~\ref{Fig5}. Also shown are the rate performance of strategies in which both relays perform DF \cite[Sec. IV.C]{KramerGastparGupta2005} or NNC, as well as  the cut-set bound. Note that, for this network, condition \eqnref{full}   (and hence \eqnref{remark1})  is not always satisfied.  The quantization noise variance $\hat N_3$ is optimized numerically. We observe that the proposed scheme reduces the gap to the cut-set bound and outperforms the other strategies for the considered topology. The increasing gap between the proposed scheme and DF-SNNC without rate splitting illustrates the gain of the former: without rate-splitting, the mixed strategy  is limited by the rate to the DF relay that suffers interference from the NNC relay. 
\section{Discussion and Future Work} \label{Conclusion}
We presented a new relaying strategy that combines SNNC with superposition coding.
In the proposed encoding scheme, superposition coding is used to encode the quantization index sent by a SNNC relay, in order to facilitate partial decoding of quantization index at relays performing DF. 
We demonstrated that this relaying strategy can bring rate gains compared to NNC, DF, as well as mixed  strategies in which some of the nodes use DF and others SNNC.  
Analysis of the proposed SNNC-RS encoding scheme in larger networks where mutliple nodes can benefit from this scheme and therefore further increase the achievable rate, is the topic of our future work.

\section{Acknowledgment} \label{Ack}
The authors would like to thank Song-Nam Hong for insightful discussion regarding  performance results.
\section {Appendix: Proof of Theorem \thmref{MainResult}}
\begin{IEEEproof} 

\begin{table*}
\caption{DF and SNNC-RS for the Two-relay Channel}
\centering
\begin{tabular}{c | c c c c}
\hline
Block & 1 & 2 & 3 & 4 \\
\hline
$X_1$       & $x_{11}(1,w_1)$  & $x_{12}(w_1,w_2)$             & $x_{13}(w_2,w_3)$             & $x_{14}(w_3,1)$ \\ 
$X_2$       & $x_{21}(1)$        & $x_{22}(w_1)$                   & $x_{23}(w_2)$                    & $x_{24}(w_3)$\\ 
$X_{30}$   & $x_{(30)1}(1)$    & $x_{(30)2}(u_{01})$           & $x_{(30)3}(u_{02})$            & $x_{(30)4}(u_{03})$  \\ 
$X_{31}$   & $x_{(31)1}(1,1)$ & $x_{(31)2}(u_{11}, u_{01})$ & $x_{(31)3}(u_{12}, u_{02})$ & $x_{(31)4}(u_{13}, u_{03})$\\ 
$\hat Y_3$ & $\hat y_{31}(u_{11}, u_{01} | 1,1)$  & $\hat y_{32}(u_{12}, u_{02} | u_{11}, u_{01})$  & $\hat y_{33}(u_{13}, u_{03} | u_{12}, u_{02})$  & $\hat y_{34}(u_{14}, u_{04} | u_{13}, u_{03})$\\
\hline
\end{tabular}
\end{table*}
Fix a distribution $p(x_1 x_2)p(x_3)p(\hat y_3|x_3 y_3)$.
The source message $w$ containing $nBR$ bits is split into messages $w_1, w_2, \ldots, w_B$ of $nR$ bits each. 
Transmission is performed over $B+2$ blocks. The last block is used to transmit index $u_B$ from node 3 to the destination.
Table I shows the encoding at the nodes for $B = 3$ blocks.

{\it Codebook generation:}
For each block $b, b=1, \ldots, B+1$ generate $2^{nR}$ codewords $\xv_{2b}(w_b)$  $w_b = 1, \ldots, 2^{nR}$ according to $\prod_{i=1}^n P_{X_2}(x_{2bi})$.
For each $\xv_{2b}(w_{b})$, generate $2^{nR}$ codewords $\xv_{1b}(w_b, w'_b)$, $w'_b = 1, \ldots, 2^{nR}$ according to $P_{X_1|X_2}$.
Generate $2^{nR_{30}}$ codewords $\xv_{(30)b}(u_{0b})$, $u_{0b} = 1, \ldots, 2^{nR_{30}}$ according to $\prod_{i=1}^n P_{X_{30}}(x_{(30)bi})$.
For each $\xv_{(30)b}(u_{0b})$, generate $2^{nR_{31}}$ codewords $\xv_{(31)b}(u_{1b}, u_{0b})$, $u_{1b} = 1, \ldots, 2^{nR_{31}}$ according to $\prod_{i=1}^n P_{X_{31}|X_{30}}(x_{(31)bi}|x_{(30)bi})$.
We denote $u_b=(u_{0b},u_{1b})$ and $R_3 = R_{30}+R_{31}.$
For each  $(\xv_{30b}(u_{0b}),  \xv_{31b}(u_{1b},u_{0b}))$, generate $2^{n R_3}$ codewords $\hat \yv_3(u_b|u_{0(b-1)}, u_{1(b-1)})$ $u_b = 1, \ldots, 2^{nR_3}$ according to $\prod_{i=1}^n P_{\hat Y_3|X_{30} X_{31}} (\hat y_{3i}|x_{30i} (u_{0(b-1)}) x_{31i}(u_{0(b-1)}, u_{1(b-1)}))$.

{\it Encoding:}
{\it The source:} In each block $b=1, \ldots, B+1$, the source transmits $\xv_{1b}(w_{b-1}, w_b)$. By convention, we set $w_0=1, w_{B+1} = 1.$

{\it Relay 1:} 
Relay 1 decodes $(w_b, u_b)$ using sliding window decoding, as described in the decoding part.
That allows relay 1 to decode $w_b$ at the end of block $b$.
In block $b+1$, relay 1 transmits $\xv_{2(b+1)}(w_b)$.

{\it Relay $2$:}
After receiving $\yv_{3b}$, at the end of block $b$, relay 2 tries to find an index $u_{0b}$ such that
\begin{align}
& ( \hat \yv_3(u_{0b}, u_{1b}|u_{0(b-1)}, u_{1(b-1)}), \xv_{(30)b}(u_{0(b-1)}), \nonumber \\
& \xv_{(31)b}(u_{0(b-1)}, u_{1(b-1)}), \yv_{3b})\in T_{\epsilon}^n(P_{X_{30} X_{31} \hat Y_3 Y_3 }) \eqnlabel{quantization}
\end{align}
where $u_{00}=1, u_{01}=1$ by convention.
 In each block b, $b=1, \ldots, B+2$, relay 2  transmits $ \xv_{31b}(u_{0(b-1)},u_{1(b-1)})$.

{\it Decoding.} 

{\it Relay 1:}
At the end of block $b$, relay 1 decodes $(w_b, u_b)$ using sliding window decoding, as follows.
 At the end of block $b=2, \ldots, B+1$, relay $1$ determines $(\hat w_b, \hat u_{0(b-1)})$ s.t.
\begin{align}
&(  \xv_{1b}(w_{b-1}, \hat w_b), \xv_{2b}(w_{b-1}) \xv_{(30)b}(\hat u_{0(b-1)}), \yv_{2b})  \nonumber \\
& \in T_{\epsilon}^n(P_{X_1 X_2 X_{30}  Y_2}) \eqnlabel{relaydecoder}
\end{align}
where $u_{00}=1$ by convention and assuming that, in the previous block, relay 1 had correctly decoded $w_{b-1}$, i.e., $\tilde w_{b-1} = w_{b-1}$.
Note that relay $1$ knows $w_0=0$ which allows relay 1 to start the sliding window decoding. 

{\it Destination:} Destination uses backward decoding \cite{KramerGastparGupta2005} as follows.

At the end of block $B+2$, node $4$ has reliably decoded $(u_{0(B+1)},u_{1(B+1)})$ that was transmitted in the  block $B+2$ from node $3$.
At the end of block $B+1$, node $4$ tries to find $(\hat w_B, \hat u_{0B}, \hat u_{1B})$ such that
\begin{align}
&\left (\xv_{1(B+1)}(\hat w_B, 1), \xv_{2(B+1)}(\hat w_B), \right. \nonumber \\
 & \hat \yv_{3(B+1)}(u_{1(B+1)} u_{0(B+1)}|\hat u_{1B} \hat u_{0B}),  \xv_{(30)(B+1)}(\hat u_{0B}), \nonumber \\
& \left. \xv_{(31)(B+1)}(\hat u_{0B}, \hat u_{1B}), \yv_{4(B+1)} \right) \in T^n_{\epsilon}(P_{X_1 X_2 X_{30} X_{31} \hat Y_3 Y_4}). \eqnlabel{decoding}
\end{align}
For blocks $b=B, B-1, \ldots, 2$, node $4$ tries to find $(\hat w_{b-1}, \hat u_{0(b-1)}, \hat u_{1(b-1)})$ such that
\begin{align}
&\left( \xv_{1b}(\hat w_{b-1}, w_b), \xv_{2b}(\hat w_{b-1}), \hat \yv_{3b}(u_{1b} u_{0b}|\hat u_{1(b-1)} \hat u_{0(b-1)}), \right. \nonumber \\
 & \left. \xv_{(30)b}(\hat u_{0(b-1)}),  \xv_{(31)b}(\hat u_{0(b-1)}, \hat u_{1(b-1)}), \yv_{4b} \right) \nonumber \\
& \qquad \qquad \in T^n_{\epsilon}(P_{X_1 X_2 X_{30} X_{31} \hat Y_3 Y_4})
\end{align}
where $(w_b, u_{0b}, u_{1b})$ have already been reliably decoded in the previous block $b+1$.

{\it Error Probability:}

Assume without loss of generality that $w_b=1$ and $u_{0b}=1, u_{1b}=1$.
We denote the event defined by \eqnref{relaydecoder} with $E_{1b}(w_{b-1}, w_b, u_{b-1})$.

The decoder at relay 1 makes an error if one of the following error events occur:
\begin{align}
&{\cal E}_{r1} =  E_{1b}^c(1,1,1)  \nonumber \\
&{\cal E}_{r2} = \bigcup_{w_b \neq 1} E_{1b}(1,w_b,1) \nonumber \\
&{\cal E}_{r3} = \bigcup_{(w_b, u_{b-1}) \neq (1,1)} E_{1b}(1,w_b,1) \nonumber
\end{align} 

Note that the event
\begin{align}
&{\cal E}'_{r} = \bigcup_{u_{b-1} \neq 1} E_{1b}(1,1,u_{b-1}) \nonumber \\
\end{align} 
is not an error event because relay 1 does not need to reliably decode $u_b$.

The error event at relay 1 is given by ${\cal E}_r = \cup_{i=1}^3 {\cal E}_{ri}$.
By the union bound, the probability of error at relay 1 is  given by
\begin{equation}
{\cal E}_r  \le \sum_{i=1}^3 P( {\cal E}_{ri})
\end{equation}

At the destination node, the decoder decodes three indexes as specified by \eqnref{decoding}, resulting in eight possible error events that can be defined similarly to the case of node $2$. 

Finally, quantization at relay $2$,  \eqnref{quantization},  demands that
\begin{equation}
R_{30}+ R_{31} > I(\hat Y_3; Y_3|X_3).   \eqnlabel{ratedistortion}
\end{equation}

To quarantee that the decoding error at two receivers becomes small as $n$ gets large, by using the standard procedure to bound the probability of error events \cite{Cover}, it can be shown that the following rate bounds need to be satisfied:
\begin{align}
& R < I(X_1;Y_2|X_2 X_{30}) \eqnlabel{JD1} \\
&R < I(\bar X; \hat Y_3 Y_4|X_3) \eqnlabel{JD2} \\
&  R_{30}+ R_{31} < I(\hat Y_3; \bar X Y_4|X_3) +  I(X_3; Y_4|\bar X)  \eqnlabel{JD3} \\
& R +  R_{30}+  R_{31}  <  I(\hat Y_3; \bar X Y_4|X_3) +  I(\bar X X_3; Y_4)  \eqnlabel{JD4} \\
& R_{31} < I(\hat Y_3; \bar X Y_4|X_3)  + I (X_{31}; Y_4|\bar X X_{30})   \eqnlabel{JD5} \\
& R + R_{31} < I(\hat Y_3; \bar X Y_4|X_3)  + I (\bar X X_{31}; Y_4| X_{30})  \eqnlabel{JD6} \\
&R + R_{30} < I(X_1 X_{30};Y_2|X_2 ) \eqnlabel {JD9}\\
&R_{30}+ R_{31} > I(\hat Y_3; Y_3|X_3)   \eqnlabel{JD7} 
\end{align}
for the joint distribution given by \eqnref{pmf}, where \eqnref{JD7} is added due to \eqnref{ratedistortion}.

Note that we cannot ignore error events in which message is correctly decoded, but one or both quantization indexes are not (which would allow to omit \eqnref{JD3} and \eqnref{JD5}). 
This is because correct quantization indexes are needed for backward decoding.
On the other hand, we can ignore an error event at relay $1$ in which only the quantization index part (but not the message) is decoded in error.

We perform Fourier-Motzkin elimination of $R_{30}$ to obtain 
\begin{align}
& R < I(X_1;Y_2|X_2 X_{30}) \nonumber \\
&R < I(\bar X; \hat Y_3 Y_4|X_3) \nonumber \\
&R < I (\bar X X_3; Y_4)  - I(\hat Y_3; Y_3|\bar X X_3 Y_4) \nonumber \\
&   R_{31} < I(\hat Y_3; \bar X Y_4|X_3) +  I(X_{31}; Y_4|\bar X X_{30})  \nonumber \\
& R + R_{31} < I(\hat Y_3; \bar X Y_4|X_3) +  I(\bar X X_{31}; Y_4| X_{30})  \nonumber \\
& R- R_{31} < I(X_1 X_{30}; Y_2|X_2)-I(\hat Y_3; Y_3|X_3) \nonumber \\
&I( \hat Y_3;  Y_3|\bar X X_{30} X_{31} Y_4) < I( X_{30} X_{31}; Y_4|\bar X)   \eqnlabel{conditionnotneeded1}
\end{align}
where we omitted loose inequalities.

Finally, by performing Fourier-Motzkin elimination to eliminate $R_{31}$, we obtain
\begin{align}
& R < I(X_1;Y_2|X_2 X_{30}) \eqnlabel{jd1R} \\
&R < I(\bar X; \hat Y_3 Y_4|X_{30} X_{31}) \eqnlabel{jd2R} \\
& R < I (\bar X X_{30} X_{31}; Y_4) -  I(\hat Y_3; Y_3|\bar X X_{30} X_{31} Y_4)    \eqnlabel{jd3R} \\
&R <I ( X_{31}; Y_4| \bar X X_{30}) - I(\hat Y_3; Y_3|\bar X X_{30} X_{31} Y_4)   \nonumber \\
& \qquad + I(X_1 X_{30}; Y_2|X_2) \eqnlabel{jd9R} \\
&2R <  I (\bar X X_{31}; Y_4| X_{30}) - I(\hat Y_3; Y_3|\bar X X_{30} X_{31} Y_4)  \nonumber \\
& \qquad   + I(X_{30}; Y_2|\bar X) +I(X_1;Y_2|X_2)\eqnlabel{2RR} \\
&I( \hat Y_3;  Y_3|\bar X X_{30} X_{31} Y_4) < I( X_{30} X_{31}; Y_4|\bar X)   \eqnlabel{conditionnotneededR}
\end{align}
Condition \eqnref{conditionnotneededR} is equivalent to the condition in the single-relay channel given in \cite[Eq.~11]{KramerHou2011}. 
When condition \eqnref{conditionnotneededR} is not satisfied, the bound reduces to
\begin{align}
& R < I(X_1;Y_2|X_2 X_{30}) \\
& R < I(\bar X;Y_4)
\end{align}
which can be achieved by treating the signals from the relay $2$ as noise, similarly as in \cite{YassaeeAref2011}. Therefore, condition \eqnref{conditionnotneededR} can be omitted and  we obtain rate bounds given by Thm.~\thmref{MainResult}. 
\end{IEEEproof}


\bibliographystyle{IEEEtran}

\begin{thebibliography}{1}
\providecommand{\url}[1]{#1}
\csname url@samestyle\endcsname
\providecommand{\newblock}{\relax}
\providecommand{\bibinfo}[2]{#2}
\providecommand{\BIBentrySTDinterwordspacing}{\spaceskip=0pt\relax}
\providecommand{\BIBentryALTinterwordstretchfactor}{4}
\providecommand{\BIBentryALTinterwordspacing}{\spaceskip=\fontdimen2\font plus
\BIBentryALTinterwordstretchfactor\fontdimen3\font minus
  \fontdimen4\font\relax}
\providecommand{\BIBforeignlanguage}[2]{{%
\expandafter\ifx\csname l@#1\endcsname\relax
\typeout{** WARNING: IEEEtran.bst: No hyphenation pattern has been}%
\typeout{** loaded for the language `#1'. Using the pattern for}%
\typeout{** the default language instead.}%
\else
\language=\csname l@#1\endcsname
\fi
#2}}
\providecommand{\BIBdecl}{\relax}
\BIBdecl

\bibitem{Avestimehr2011}
A.~Avestimehr, S.~N. Diggavi, and D.~Tse, ``Wireless network information flow:
  A deterministic approach,'' \emph{IEEE Trans. Inf. Theory}, vol.~57, pp.
  1872--1905, Apr. 2011.

\bibitem{LimKimElGamal2011}
S.~Lim, Y.-H. Kim, A.~{El Gamal}, and S.-Y. Chung, ``Noisy network coding,''
  \emph{IEEE Trans. Inf. Theory}, vol.~57, no.~5, pp. 3132--3152, May 2011.

\bibitem{HouKramer2013}
J.~Hou and G.~Kramer, ``Short message noisy network coding with a
  decode-forward option,'' \emph{submitted to IEEE Trans. Inf. Theory}, Aug.
  2013.

\bibitem{CoverElGamal79}
T.~Cover and A.~{El Gamal}, ``Capacity theorems for the relay channel,''
  \emph{IEEE Trans. Inf. Theory}, vol.~25, no.~5, pp. 572--584, Sep. 1979.

\bibitem{KramerHou2011}
G.~Kramer and J.~Hou, ``Short-message quantize-forward network coding,'' in
  \emph{2011 8th Int. Workshop on Multi-Carrier Systems Solutions(MC-SS),
  Herrsching, Germany,}, May 2011, pp. 1--3.

\bibitem{KramerGastparGupta2005}
G.~Kramer, M.~Gastpar, and P.~Gupta, ``Cooperative strategies and capacity
  theorems for relay networks,'' \emph{IEEE Trans. Inf. Theory}, vol.~51,
  no.~9, pp. 3037--3063, Sep. 2005.

\bibitem{Cover}
T.~Cover and J.~Thomas, \emph{Elements of Information Theory}.\hskip 1em plus
  0.5em minus 0.4em\relax John Wiley Sons, Inc., 1991.

\bibitem{MaricArxiv2014}
I.~Mari{\'c} and D.~Hui, ``Short message noisy network coding with rate
  splitting,'' \emph{http://arxiv.org/abs/1404.0061}, Apr. 2014.

\bibitem{YassaeeAref2011}
M.~Yassaee and M.~Aref, ``Slepian-{W}olf coding over cooperative networks,''
  \emph{IEEE Trans. Inf. Theory}, vol.~57, no.~6, pp. 3462--3482, 2011.

\end{thebibliography}
\end{document}